# A Novel Recommendation System to Match College Events and Groups to Students

K Qazanfari[1,*], A Youssef[1], K Keane[2] and J Nelson[2]

[1] Department of computer science, George Washington University, Washington, DC, USA 20052

[2] Promantus Inc. Cary, NC, USA 27511

*Kazemmit@gwu.edu

**Abstract.** With the recent increase in data online, discovering meaningful opportunities can be time-consuming and complicated for many individuals. To overcome this data overload challenge, we present a novel text-content-based recommender system as a valuable tool to predict user interests. To that end, we develop a specific procedure to create user models and item feature-vectors, where items are described in free text. The user model is generated by soliciting from a user a few keywords and expanding those keywords into a list of weighted near-synonyms. The item feature-vectors are generated from the textual descriptions of the items, using *modified tf-idf* values of the users' keywords and their near-synonyms. Once the users are modeled and the items are abstracted into feature vectors, the system returns the maximum-similarity items as recommendations to that user. Our experimental evaluation shows that our method of creating the user models and item feature-vectors resulted in higher precision and accuracy in comparison to well-known feature-vector-generating methods like Glove and Word2Vec. It also shows that stemming and the use of a modified version of tf-idf increase the accuracy and precision by 2% and 3%, respectively, compared to non-stemming and the standard tf-idf definition. Moreover, the evaluation results show that updating the user model from usage histories improves the precision and accuracy of the system. This recommender system has been developed as part of the Agnes application, which runs on iOS and Android platforms and is accessible through the Agnes website.

## 1. Introduction

With a world of exponential growth of online information, finding useful data can be surprisingly difficult [1]: there are too many options for users to assess, sort through, and find the ones most relevant to their interests. To remedy the information overload, recommender systems could be a valuable tool for online users [1].

Recommender Systems or Recommendation Systems (RSs) are a subclass of information filtering systems that collect information on the preferences of users for a set of items such as movies, songs, books [2, 3]. RS might use different user information such as age and nationality, or social information such as followers, followed, tweets, and posts [2]. RSs compare the user's information to available items against reference characteristics in order to find recommendations [1]. Generally, a recommendation system compares the user profile to some reference attributes to predict the 'rating' that a user would give to an item that has not been seen by the user [1].

Generally, the recommender systems are classified into the following three categories:

- Collaborative recommendation systems: These approaches recommend the items based on the similarity of tastes and preferences between users using a utility matrix [4, 5].
- Content-based recommendation systems: These approaches recommend the items based on the description of the item and a profile of the user's preference and history [6-10].
- Hybrid recommendation systems: These approaches combine collaborative and content-based methods to recommend some items to the users [11, 12].

Since the content-based systems are designed mostly to recommend text-based items [6], the rest of this subsection will focus on reviewing of some text-content-based recommender systems. Recently, considerable research has been done in the field of text-based recommendation systems. Gu et al. [7] proposed a method to measure the similarity between two pieces of text using cosine similarity of the two bags of words, where each word is weighted by its tf-idf, and they applied this method on LinkedIn job data. A similar method has been proposed by Philip et al. [8] but for digital libraries. Also, Lak et al. [9] proposed an article recommendation system that uses tf-idf and word2vec for converting each article to a feature-vector. Some more comprehensive notion of document relevance than bags-of-words are proposed in [13, 14]; they extract a wide variety of content-based features to characterize non-linguistic aspects of the audio such as speaker, language, gender, and environment. In [13], authors use Glove for feature extraction and using them into CNN.

Although hybrid recommendation systems might provide higher performance, the collaborative RS component needs sufficient rated items to create the utility matrix. However, since often there is not enough rated items in exclusive communities for creating the utility matrix, this paper proposes a novel text-content-based recommendation systems for recommending some items (e.g. college events) to users (e.g. students). By considering the unstructured text content of the items, we convert the content of these items to some meaningful feature-vectors by computing a modified version of tf-idf on some specific pre-processed words. Finally, we find the similarity between feature-vectors using cosine measurement. The details of the proposed method are presented in the next section.

The rest of this paper is organized as follows: The proposed recommendation system will be explained in Section 2. In Section 3, experiments will be presented and analyzed. The paper concludes in Section 4 with a discussion of the results achieved and some suggestions of lines of future work in the field.

**2. Proposed recommendation system**

In this section, we propose a novel content-based recommendation system to increase the user engagement in exclusive communities. In the content-based recommender systems, the recommendation decisions are made based on the item content and the user content. In our method, items are activities that take place at a specific time and specific place, and described in free text, like the following item describing an event in a college community:

**Event Title**, e.g., "Smithsonian American Art Museum Highlights Tour"
**Event Description**, e.g., "This highlights tour of the museum's collection may include highlights of the temporary exhibitions. Location: F Street Lobby."
**Event Tags**, e.g., Art, Museum

Also, the user content includes a few keywords that are solicited from a user and that represent his/her interests, hobbies, and needs such as Painting, Cooking, Fitness and Traveling. The following gives the main processes of our recommendation system:
A. Creating the initial user model for each new user based on keywords from the user.
B. Updating the user model based on user history (negative and positive items).
C. Recommending some new matched items to the user based on his/her model.

Each of the above processes will be explained and illustrated in the following sections.

*2.1. Creating the initial user model*

Based on the targeted community, each user has some interests, and the intensity of interests should be quantified. For example, if a user is very interested in "sport", we model it as {sport: 5}, or if a user is not interested in "sport", we model it as {sport: -2}, so these numbers reflect the intensity of user interest.

In this section, we introduce the process of creating the initial user model; an array of {word: weight} pairs. This array will be created based on the keywords provided by the user.

The user model includes two types of word: the keywords that are provided by the user directly; and words that are the *near-synonyms* of those keywords. The near-synonyms of a keyword are the words that are synonymous or similar (in meaning) to the keyword. The number of near-synonyms of a keyword is a parameter ($S$) that is set before creating the initial user model.

Regarding the keyword weight, we assign initially a constant $W_{max}$. The weight of a near-synonym of a keyword ($W_{near-synonym}$), is a number between 0 (exclusive) and 1 (inclusive), where higher weights signify more similarity. These near-synonym weights are computed by the near-synonym generator such as Glove. For example, if a user provides "sport" and "technology" as keywords, and we take $S = 5$ and $W_{max} = 2$, then the process of creating the initial user model is:

1) Collecting $S=5$ near-synonyms of the "sport" and "technology" keywords as shown in Table 1:

Table 1. Near-synonyms of the "sport" and "technology" keywords.

| **near-synonyms** of "sport": | athletics | football | rowing | racing | wrestling |
|---|---|---|---|---|---|
| $W_{near-synonym}$: | 1.0 | 1.0 | 0.9 | 0.9 | 0.8 |
| **near-synonyms** of "technology": | engineering | IT | application | business | technological |
| $W_{near-synonym}$: | 0.8 | 0.8 | 0.6 | 0.5 | 0.5 |

2) Creating the initial user model by computing the union of all keywords and their near-synonyms. If there is a duplicated word, use the highest weight among all duplicates as shown in Table 2.

Table 2. A sample initial model based on "sport" and "technology" keywords.

| **word:** | sport | athletics | Football | Rowing | Racing | Wrestling | technology | engineering | IT | application | business | technological |
|---|---|---|---|---|---|---|---|---|---|---|---|---|
| **Weight:** | 2.0 | 1.0 | 1.0 | 0.9 | 0.9 | 0.8 | 2.0 | 0.8 | 0.8 | 0.6 | 0.5 | 0.5 |

*2.2. Refining the user model and finding top recommended items for users*

In this section, we will explain how to refine the user model and find the top *N* matched items for each user. From the viewpoint of having useful information, we suppose that the content of each item has different textual parts; we call each part a *data field*. For example, an event item in a college community might have three different data fields as Event Title, Event Description and Event Tags, and each data field has a different amount of useful information, which is called the *significance weight* of the data field. The significance weights are parameters in our system, set at configuration time of the RS.

To find the top *N* matched items, the following steps are performed:
1) Refine the user model into a model $User_{RM}$ as follows:
    a. Calculate the lemmatized and/or stemmed versions of the initial/updated user model and call the resulting model $Model_L$ and/or $Model_S$
    b. For each keyword or near-synonym, remove the last letter of the word if the word ends with 'e', 'y' or 'i' based on the following conditions:

i. After lemmatizing, if a word ends with 'e' or 'y', remove the last letter
ii. After stemming, if a word ends with 'i', remove the last letter
c. We allow lemmas and stems of length at least 3 letters. Therefore, after performing steps 1.a and 1.b, remove all words of length 2 or less from $Model_L$ and $Model_S$
d. If we only use the lemmatized version of the user model, set $User_{RM} = Model_L$
e. If we only use the stemmed version of the user model, set $User_{RM} = Model_S$
f. If we use both lemmatized and stemmed versions, set $User_{RM}$ to be the union of $Model_L$ and $Model_S$
g. In all the above cases, if we have a duplicated word, we should keep only one copy of it and use the highest weight among all duplicates as the word weight
2) Let $U_v$ be the weight vector in e $User_{RM}$
3) For each item $i$, calculate the similarity $W_i$ between the item feature-vector $S_i$ and user weight vector $U_v$ as follows:
a. For each data field $j$ with significance weight $W_{df(j)}$:
i. $S_{df(j)} :=$ the tf-idf for the $j^{th}$ data field of the item for the words that we have in the $User_{RM}$;
b. Calculate the item feature-vector $S_i$ by weighted summation over all data fields:
$$S_i = \sum_{j=1}^{n=number\ of\ data\ fields} S_{df(j)} * W_{df(j)} \qquad (1)$$
c. Calculate $W_i$ as the similarity between $S_i$ and $U_v$; (similarity measures will be addressed later)
4) Sort all items based on calculated $W$'s in decreasing order.
5) Select top $N$ items.

To make the above processes clear, we illustrate them with an example: Consider the initial user model as shown in Table 2, and using stemming for Step 1.a. Also, suppose that the target community has items with three data fields.

In this example, we use the stemmed version of the user model for creating $User_{RM}$. Based on the mentioned user model, $Model_S$ will be changed as shown in Table 3. In this table, for each step, the cells which were affected by this step are colored in gray.

Table 3. Refining the user model based on the initial model of Table 2.

| User model words: | sport | athletics | football | rowing | racing | wrestling | technology | engineering | IT | application | business | technological |
|---|---|---|---|---|---|---|---|---|---|---|---|---|
| Step 1.a. | sport | athlet | football | row | race | wrestl | technolog | engin | IT | applic | busi | technolog |
| Step 1.b. | sport | athlet | football | row | race | wrestl | technolog | engin | IT | applic | Bus | technolog |
| Step 1.c. | sport | athlet | football | row | rac | wrestl | technolo | engin |  | applic | bus | technolo |

| Step 1.g. | sport | athlet | fotbal | row | rac | wrestl | technolog | engin | | applic | bus | |
|---|---|---|---|---|---|---|---|---|---|---|---|---|
| **Weight** | 2.0 | 1.0 | 1.0 | 0.9 | 0.9 | 0.8 | 2.0 | 0.8 | 0.8 | 0.6 | 0.5 | 0.5 |

Finally, Table 4 shows the refined model $User_{RM}$.

**Table 4**. Refined model on the initial model of Table 2.

| $User_{RM}$ words: | sport | athlet | football | row | race | wrestl | technolog | engin | applic | bus |
|---|---|---|---|---|---|---|---|---|---|---|
| $User_{RM}$ words vector ($U_v$): | 2.0 | 1.0 | 1.0 | 0.9 | 0.9 | 0.8 | 2.0 | 0.8 | 0.6 | 0.5 |

Now, suppose we have 3 different college events as items. The first one is about honoring a football team, the second item is about course registration, and the last one is about using technology in health. The calculated tf-idf of each mentioned item for $User_{RM}$ words are as mentioned in Table 5.

**Table 5.** Calculated tf-idf for $User_{RM}$ words for three items.

| $User_{RM}$ words: | | sport | athlet | footbal | row | race | wrestl | technolog | engin | applic | bus |
|---|---|---|---|---|---|---|---|---|---|---|---|
| item 1 | $S_{df(1)}$: | 0 | 1.70 | 1.0 | 0 | 0 | 0 | 0 | 0 | 0 | 0 |
| | $S_{df(2)}$: | 2.60 | 3.40 | 2.0 | 0 | 0 | 0 | 0 | 0 | 0 | 0 |
| | $S_{df(3)}$: | 1.30 | 0 | 1.0 | 0 | 0 | 0 | 0 | 0 | 0 | 0 |
| item 2 | $S_{df(1)}$: | 0 | 0 | 0 | 0 | 0 | 0 | 0 | 0.50 | 1.30 | 0 |
| | $S_{df(2)}$: | 0 | 0 | 0 | 0 | 0 | 0 | 0 | 0 | 0 | 0 |
| | $S_{df(3)}$: | 0 | 0 | 0 | 0 | 0 | 0 | 0 | 0 | 0 | 0 |
| item 3 | $S_{df(1)}$: | 0 | 0 | 0 | 0 | 0 | 0 | 0.82 | 0 | 0 | 0 |
| | $S_{df(2)}$: | 1.30 | 0 | 0 | 0 | 0 | 0 | 2.46 | 0 | 0 | 0 |
| | $S_{df(3)}$: | 0 | 0 | 0 | 0 | 0 | 0 | 0 | 0.50 | 0 | 0 |

If we consider $W_{df(1)} = 1.0$, $W_{df(2)} = 1.2$ and $W_{df(3)} = 0.8$, then the item vector $S_i$ of these three items are given in Table 6.

**Table 6.** Calculated item vector $S_i$ for above three items.

| Final words | sport | athlet | Football | Row | Race | Wrestl | technolog | engin | applic | bus |
|---|---|---|---|---|---|---|---|---|---|---|
| $S_1$ | 3.64 | 4.42 | 3.8 | 0 | 0 | 0 | 0 | 0 | 0 | 0 |
| $S_2$ | 0 | 0 | 0 | 0 | 0 | 0 | 0 | 0.5 | 1.3 | 0 |
| $S_3$ | 1.04 | 0 | 0 | 0 | 0 | 0 | 2.788 | 0.6 | 0 | 0 |

Now, we can calculate the similarity between each item vector $S_i$ and user vector $U_v$. To calculate the similarity, we can use any similarity measure such as the cosine similarity that makes more sense for our problem. If $a = (a_1, a_2, \ldots a_n)$ and $b = (b_1, b_2, \ldots b_n)$ are two n-dimensional vectors, then the cosine similarity between these two vectors, $\cosine(\theta)$, is represented using a dot product and magnitude as:

$$Similarity(a,b) = cosine(\theta) = \frac{a.b}{\|a\|\|b\|} = \frac{\sum_{i=1}^{n} a_i b_i}{(\sum_{i=1}^{n} a_i^2)^{1/2} (\sum_{i=1}^{n} b_i^2)^{1/2}} \tag{2}$$

Based on above formula, the similarity between $U_v$ and each $S_i$ will be:

$W_1 = Similarity(U_v, S_1) = 0.61$; $W_2 = Similarity(U_v, S_2) = 0.23$; $W_3 = Similarity(U_v, S_3) = 0.73$

Finally, if the number of desired recommendations is $N = 2$, then we will recommend to the user the two items with the top $W_i$ values, namely, item 3 and item 1.

*2.3. Updating the user model (Learning process)*

In this section, we introduce a method for updating the user model based on his/her history. Each user has some history data, including some negative and positive samples, where the negative samples are the items that the system recommended but the user did not like, and the positive samples are the items that the user liked. The process of updating the user model involves two major steps:

1) Updating the user model words.
2) Updating the user model weights.

*2.3.1. Updating the user model words.* In the previous section, we introduced how to create the initial user model for each user. In this section, we explain how our method updates the user model based on user history. Consider the sample initial user model as shown in Table 2.

For updating the user model words, we just need to add whatever words in all other users' models that do not exist in the current user model. We also consider weight=0 for any newly added word. For example, consider the above-mentioned user model and some other words from other users' models including "art", "dance", "artwork", "painting", "sculpture" and "artistry". Then, Table 7 shows the new user model.

**Table 7.** Updated model of the initial user model Table 2.

| Word: | sport | athletics | football | rowing | racing | wrestling | technology | engineering | IT | application | business | technological | art | dance | artwork | painting | sculpture | artistry |
|---|---|---|---|---|---|---|---|---|---|---|---|---|---|---|---|---|---|---|
| Weight: | 2.0 | 1.0 | 1.0 | 0.9 | 0.9 | 0.8 | 2.0 | 0.8 | 0.8 | 0.6 | 0.5 | 0.5 | 2.0 | 0 | 0 | 0 | 0 | 0 |

In the next section, we describe how we can update the weights of the created model.

*2.3.2. Updating the user model weights.* The process of updating the user model weights is:

1) Let the current user model be $M_c$, which is the output of the previous step (Updating the user model words).
2) For each rated item $Q$:
   a. Calculate the model of $Q$ as $M_Q$ (explained in the next paragraph)
   b. If $Q$ was labeled as a negative sample, then $M_c = M_c - \alpha * M_Q$
   c. If $Q$ was labeled as a positive sample, then $M_c = M_c + \alpha * M_Q$

Now, we explain how we can calculate the item model $M_Q$. This process is similar to the process of finding top items. However, there are some differences. Consider a positive or negative item $Q$:

   a. For each data field $j$ with significance weight $W_{df(j)}$:
      i. $S_{df(j)} :=$ the tf-idf for the $j^{th}$ data field of item $Q$ for the words that we have in the updated user model;
   b. Calculate the item vector $M_Q$ by weighted summation over all the $n$ data fields:

$$M_Q = \sum_{j=1}^{n}[S_{df(j)} * W_{df(j)}] \qquad (3)$$

## 3. Performance Evaluation

In this section, we evaluate experimentally the performance of our content-based recommendation system and various design decision choices. To do so, the college community is selected as a good sample of data-overwhelmed exclusive community to be used as a testbed. Therefore, our goal is to recommend some college events to students. All data that we used in these experiments is collected from the George Washington University website. However, we needed some of this data to be labeled by humans so we could compare the result of our method with labeled data.

To obtain this labeled data, we provided 10 different workbooks. Each workbook had 30 events. Then we gave workbooks to ten different users and asked them to provide:

1) Some keywords that indicate their hobbies, activities, skills, and issues that they might be interested in.
2) 10 events out of the 30 provided, that are more interesting to them.

### 3.1. Difference between stemming and lemmatizing

As we mentioned before, words might be in plural forms or even in other forms. To convert all forms of a word to the same word, we can use stemming or lemmatization. For a single input word, these two processes might result in different forms of the word. For example, consider word "technology"; the stemming process converts it to "technolog" and the lemmatizing process converts it to "technology".

In the first experiment, we are going to compare the following four alternatives: Original words, Stemmed words, Lemmatized words and, Union of Stemmed and lemmatized.

Note that the process of removing 'e', 'y' and 'i' letters from the end of the words is not performed in this experiment, but it will be carried out in the second experiment. Table 8 shows the results of matching for the four above situations. As you can see, using stemming produces higher precision and accuracy.

**Table 8.** Evaluation of our RS for when different versions of the words are used in calculations.

| Method | Precision | Accuracy |
|---|---|---|
| Original words | 0.66 | 0.77 |
| Stemmed words | 0.73 | 0.82 |
| Lemmatized words | 0.7 | 0.80 |
| Union of Stemmed and Lemmatized words | 0.7 | 0.80 |

### 3.2. Difference between keeping or removing 'e', 'y' and 'i' letters

In the course of this research, we found that removing the letters 'e' and 'y' from the end of lemmatized words could improve the efficiency of the method. The same improvement would be happened if we remove letter 'i' from the end of stemmed words. In the next experiment, we compare the performance of our method for when 'e', 'y' and 'i' are removed from the end of stemmed words vs. when they are not removed.

**Table 9.** Evaluation of our RS for when we use the process of removing 'e', 'y' and 'i' endings and for when we do not remove them

| | Precision | Accuracy |
|---|---|---|
| Removing 'e' and 'y' (after lemmatizing) | 0.72 | 0.81 |
| Keeping 'e' and 'y' (after lemmatizing) | 0.7 | 0.80 |
| Removing 'i' (after stemming) | 0.75 | 0.83 |
| Keeping 'i' (after stemming) | 0.73 | 0.82 |

As Table 9 shows, based on which version of the words (stemming or lemmatizing) is used in the previous step, removing 'e', 'y' and 'i' improves precision and accuracy of our method. Therefore, for the rest of the experiments, stemming will be used, and the letter 'i' will be removed from the end of the stemmed versions of words.

*3.3. Difference between binary calculation or frequency calculation*

As mentioned before, we used tf-idf for creating the user and item vectors. Actually, the tf-idf metric has two parts: term frequency ($tf$) and inverse document frequency ($idf$): *tf-idf* = $tf * idf$. In some situation, it's better to binarize $tf$, that is, set tf = *1 if $tf > 0$,* and keep $tf = 0$ otherwise:

$$\text{tf-idf} = \begin{cases} idf & if \quad tf > 0 \\ 0 & if \quad tf = 0 \end{cases} \tag{4}$$

In this experiment, we want to evaluate our method for two alternatives:
- Frequency calculation: Calculating tf-idf based on its common definition.
- Binary calculation: Calculating tf-idf based on the binarized tf.

As you can see from the Table 10, the modified tf-idf method results in better precision and accuracy. We will use binary tf calculation in the remainder of the experiments.

**Table 10**. Evaluation of our RS for when the original definition of tf-idf is used and when the binarized tf is used.

|  | Precision | Accuracy |
|---|---|---|
| Frequency calculation | 0.75 | 0.83 |
| Binary calculation | 0.78 | 0.85 |

*3.4. Difference between different vector similarity measures*

As we mentioned before, we should use a good similarity measure for computing the similarity between user vector and item vector. In the next experiment, we evaluate our method based on using different similarity measures. We compare four well-known similarity measures, namely, dot-product, cosine, Euclidean, and Manhattan. Table 11 shows the performance results.

**Table 11.** Evaluation of our RS for when we use different similarity measures.

| Method | Precision | Accuracy | Method | Precision | Accuracy |
|---|---|---|---|---|---|
| Euclidean | 0.62 | 0.75 | Dot product | 0.71 | 0.81 |
| Manhattan | 0.59 | 0.73 | Cosine | 0.78 | 0.85 |

As you can see from Table 11, the cosine similarity measure resulted in the highest precision and accuracy. Therefore, we will use this measure for calculating the similarity between two vectors.

*3.5. Comparison of our method with some other methods*

In this experiment, we compare our method with the following three methods of generating a feature-vector from text content.
1) TF/IDF,
2) *Word2vec* using Google pre-trained model [15]
3) *The glove method* using [Wikipedia 2014 + Gigaword 5] model [16].

Before presenting the results of this experiment, we explain how the last two above methods finds the top *N* recommended items for users:
- word2vec Google trained model: Word2vec generates a vector for each word in an item text. The vectors of all the words in the text are added to create the feature of the item text.
- The glove method using [Wikipedia 2014 + Gigaword 5] model: It is similar to word2vec except that it uses different training algorithms.

All the above-mentioned methods including our method use the cosine measure for calculating the similarity between the user model and the item feature-vectors, and recommending the top *N* recommended items. In addition to the above methods, we compare the result of our method with UMBC phrase/sentence similarity service which is a well-known semantic textual similarity system. It gets two texts as input parameters and returns a real number as output which indicates the semantic similarity value between the two input texts. To calculate the semantic similarity between user interests and items, we feed the user keywords and the items text content as the first and second inputs of UMBC semantic similarity service respectively. Table 12 shows the result of comparison between our method and the above methods:

Table 12. Comparison between our method and some other methods.

| Method | Precision | Accuracy | Method | Precision | Accuracy |
|---|---|---|---|---|---|
| Our method | 0.78 | 0.85 | Glove | 0.68 | 0.79 |
| TF/IDF | 0.58 | 0.72 | UMBC | 0.73 | 0.82 |
| Word2vec | 0.67 | 0.78 | | | |

As you can see in table 12, although some models like the Google pre-trained model and the Glove pre-train model are strong, our method results in the best precision and accuracy. The reason for this is that these models are really general and cover so many words in the English language, whereas our problem is very specific and so our model only covers restricted vocabulary related to our problem, which leads to better results.

*3.6. Comparison of our method before and after updating the user model*
In this last experiment, we show how the updating process could improve the performance of our method. Table 13 shows the result of this experiment. As you can see, the precision and accuracy have been improved from 0.78 and 0.85 to 0.81 and 0.87 respectively.

Table 13. Evaluation of our method before and after training.

| | Precision | Accuracy |
|---|---|---|
| Before training | 0.78 | 0.85 |
| After Training | 0.81 | 0.87 |

**4. Conclusion and future work**
To overcome the data overload issue in recommendation applications, we proposed a novel text-content-based recommendation system and optimized it by considering different alternatives and different decision choices. The resulting system uses a new specific procedure to generate the user model and convert the text content of the community items to a meaningful feature-vector. The user model is generated by soliciting from a user a few keywords that represent his/her interests, and expanding those keywords into list of weighted near-synonyms. Since words might be various morphological and/or syntactical forms, we normalize all the forms of a word to the same word using stemming and/or lemmatization, and then remove 'e', 'y' and 'i' from the end of stemmed/lemmatized words. The item feature-vectors are generated from the textual descriptions of the items, using modified tf-idf values of the users' keywords and near-synonyms. Then by computing the cosine similarity between user and item feature-vectors, the items that best match the user's interests are recommended.

We tested the proposed method on a college community to recommend college events to students based on their interests. Experimental results show that stemming and removing 'i' letters from end of stemmed words increase the accuracy and precision significantly over any other combination (or lack) of stemming and lemmatization, that binarized-tf tf-idf is superior to non-binarized tf-idf, that the cosine similarity measure is superior to other similarity measures, and that model training (i.e., learning from past performance) improves the performance of the system. Also, we compared our system with some well-known methods like Glove and Word2Vec for creating the feature-vectors for text content, and found that our proposed procedure (of creating the user and item feature-vectors) resulted in highest precision (78%) and accuracy (85%) for our application among those competing methods.

As mentioned before, the hybrid recommendation systems could improve performance; however, to create the collaborative part of it, there need to be many pre-rated samples available. Currently there is no such data, but once there is, creating a collaborative or hybrid recommendation system will likely improve the performance of the proposed method. Also, creating and training a more knowledgeable model for computing more accurate near-synonyms will definitely improve the accuracy of recommendations.